\documentclass[12pt]{JHEP3}

\usepackage[centertags]{amsmath}
\usepackage{amssymb}
 \usepackage{graphicx}
\title{Thermodynamics of Plasmaballs and Plasmarings in 3+1 Dimensions}
\author{Shanthanu Bhardwaj$^{a}$, Jyotirmoy Bhattacharya$^{b}$ \\
\small{\emph{$^{a}$Department of Physics,
                      University of Chicago,}}\\
\small{\emph{Chicago, IL 60637-1434, USA.}}\\
\small{\emph{$^{b}$Department of Theoretical Physics,
                   Tata Institute of Fundamental Research,}}\\
\small{\emph{Homi Bhabha Rd, Mumbai 400005, India.}}  \\
E-mail:\ \ {\bf shanth@uchicago.edu, jyotirmoy@theory.tifr.res.in}}

\abstract{
We study localized plasma configurations in $3+1$ dimensional massive field theories obtained by Scherk-Schwarz
compactification of $4+1$ dimensional CFT to predict the thermodynamic properties of localized blackholes and 
blackrings in Scherk-Schwarz compactified $AdS_6$ using the AdS/CFT correspondence. We present an exact
solution to the relativistic Navier-Stokes equation in the thin ring limit of the fluid configuration. 
We also perform a thorough numerical analysis to obtain the thermodynamic properties of the most general
solution. Finally we compare our results with the recent proposal for the phase diagram of blackholes in six flat 
dimensions and find some similarities but other differences.
}

\preprint{TIFR/TH/09-02 \\
    \texttt{arXiv:0806.1897 [hep-th]}
}

\begin{document}

\maketitle

\section{Introduction}

Black holes in dimensions greater than four have recently drawn considerable interest 
\cite{Emparan:2007wm,Emparan:2008eg,Emparan:2001wn}. They not only provide 
testing ground of String theory (for example counting of microstates to obtain the entropy) but also are interesting in
their own right as they aid the greater understanding of General Relativity .
In this note we use the AdS/CFT correspondence to study black holes in six dimensional space-times that asymptote to
a Scherk-Schwarz compactification of $AdS_6$. Although it appears to be difficult to directly construct 
black hole backgrounds in these geometries, the AdS/CFT correspondence identifies finite energy configurations of the `deconfined gluon plasma' fluid of the dual boundary field theory with these localized  black holes permitting 
a detailed study of the spectrum of black holes and their thermodynamic properties 
in appropriate limits \cite{Aharony:2005bm}.

In more detail, the dual boundary description of gravity on a background that asymptotes to a Scherk-Schwarz 
compactification of $AdS_6$ is given by the Scherk-Schwarz compactification of a $4+1$ dimensional CFT. The effective
long distance description of this system is given by the $3+1$ fluid dynamics equations (the relativistic  Navier-Stokes equations) with the appropriate equation of state and dissipative parameters, all of which are 
directly known from gravity. In an appropriate limit, rotating bulk black holes are dual to localized rotating 
lumps of the boundary fluid that solve the Navier-Stokes equations. Such fluid configurations were investigated 
 in \cite{Lahiri:2007ae} (see \cite{Cardoso:2006sj} for a prior discussion on equilibrium fluid 
configurations and analysis of their stability in relation to the study of black holes). 
The authors of \cite{Lahiri:2007ae} constructed solutions of fluid dynamics with the 
topology of a ball and other solutions with the topology of a solid torus. They argued that, under the 
AdS/CFT correspondence described above, these solutions are dual to black objects whose horizons are topologically $S^4$ and $S^3 \otimes S^1$ respectively, i.e. black holes and black rings. 

The authors of \cite{Lahiri:2007ae}, however, left open the possibility of the 
existence of other topologically distinct solutions of fluid dynamics which would be dual to black holes 
with distinct horizon topologies. They also did not study the thermodynamic properties of the solutions 
they have constructed. Finally, the solutions of \cite{Lahiri:2007ae}, while explicit, are presented in terms of 
a function defined by a definite integral that appears difficult to control analytically.

In this paper we continue (and to an extent complete)  the analysis of \cite{Lahiri:2007ae} in three different ways. First we perform a thorough numerical scan to demonstrate that the rotating fluid solutions presented 
in \cite{Lahiri:2007ae} are the only stationary rotating solutions of the relevant Navier-Stokes equations. 
Second we determine the thermodynamic properties and the phase diagram of the solutions constructed in 
\cite{Lahiri:2007ae}. Finally we also find a particular limit (the `thin ring' limit) in which the integrals
of \cite{Lahiri:2007ae} may be explicitly evaluated. All thermodynamic properties may be evaluated 
analytically in this limit, 
allowing a better intuitive understanding of the properties of these solutions, and providing checks on 
our (numerical) determination of the thermodynamic properties of the general solution.

\begin{figure}[htp]
\centering
\includegraphics[width=0.5\textwidth]{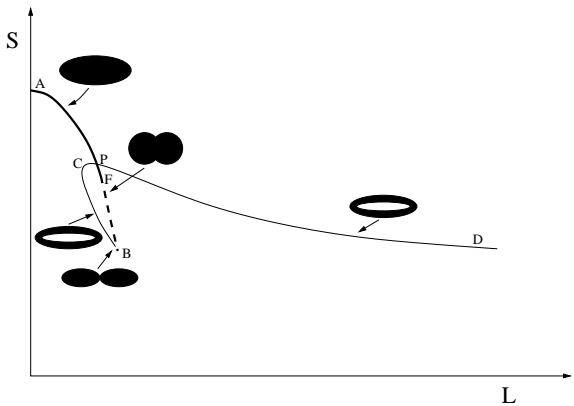}
\caption{Schematic plot of the phase diagram for the various plasma configurations 
which by AdS/CFT correspondence gives the phase structure of blackholes with various horizon topologies
in Scherk-Schwarz compactified $AdS_6$.}
\label{fig:intro}
\end{figure}

The thermodynamic properties of the ball and ring solutions of \cite{Lahiri:2007ae} turn out to be very similar 
to the properties of the analogous solutions in one lower dimension (discussed in detail in \cite{Lahiri:2007ae}).
 In fig:\ref{fig:intro} we present a plot of the entropy versus the 
angular momentum of the relevant solutions, at a fixed  
value of the energy. As is apparent from fig:\ref{fig:intro}  we find at least one rotating fluid solution 
for every value of the angular momentum. However in a particular window of angular momentum - in the range 
 $(L_B, L_C)$ - there exist three solutions which have the same energy and angular momentum. These three solutions 
may be thought of as a ball a thick ring and a thin ring respectively of rotating fluid. 
The ball solution is entropically dominant for $L<L_P$ while the thin ring dominates for $L>L_P$. At angular momentum $L_P$ (which lies in the range $(L_B, L_C)$) the system (in the microcanonical ensemble) consequently
undergoes a `first order phase transition' from the ball to the ring.  It follows that the dual gravitational
system must exhibit a dual phase transition from a black hole to a black ring at the same angular mumomentum.
 
In \S \ref{compar} we explain that the phase diagram depicted in fig:\ref{fig:intro} has some similarities, but several qualitative points of difference from a conjectured phase diagram for the solution space of rotating black holes and black rings in 6 dimensional
flat spacetime. This suggests that the properties of black holes and black rings in 
6 dimensional $AdS$ space are rather different from those of the corresponding objects in flat six dimensional space. 
This is a bit of a surprise, as black holes and rings in Scherk-Schwarz compactified $AdS_5$ appear to have properties
that are qualitatively similar to their flat space counterparts \cite{Emparan:2001wn,Lahiri:2007ae}.

The rest of this paper is organized as follows. 
In \S \ref{rev} below we review the results of \cite{Lahiri:2007ae}. 
In \S \ref{numscan} we describe the numerical scan that has convinced us that there exist no further solutions 
to the equations of  \cite{Lahiri:2007ae} in addition to the ball and ring solutions presented in that paper. 
In \S \ref{analyt} we analytically determine the ring solution and its thermodynamic properties in the `thin ring' 
limit. We also study the properties of the ball and ring solutions at the point in moduli space where they meet; 
the point at which the ball just squeeze off into a ring. In \S \ref{num} we present a thorough numerical analysis 
of the thermodynamic properties of the ball and ring solutions, and the resultant phase diagram of the system. 
in \S \ref{compar} we compare this phase diagram to the conjectured phase diagram for spinning black 
holes and black rings in flat space, presented by Emparan and collaborators \cite{Emparan:2007wm}.

\section{Review} \label{rev}

Here we briefly summarize the results of \cite{Lahiri:2007ae} as we intend to study the thermodynamic properties 
of the solutions already presented there. We also borrow our notations and conventions
from therein.
For the study of $3+1$ dimensional plasma configurations we 
work with the coordinates $(t,r,\phi,z)$ and the metric we consider
is,
\begin{equation}
ds^2 = -dt^2 + dr^2 + r^2 d\phi^2 + dz^2.
\end{equation}
Considering the axis of rotation (whenever applicable) to be the z-axis we assume that the surface of the 
plasma configuration is given by the surface $z-h(r)=0$, with $h(r)$ being a height function. The equation 
of state for the fluid under
consideration is,
\begin{equation}
F = V (\rho_0 - \lambda \mathcal{T}^5),
\end{equation}
where $F$  and $\mathcal{T}$ are the free energy and temperature respectively, $\rho_0$ denotes the vacuum energy density
and $\lambda$ being some arbitrary constant (note that $\lambda$ here is {\em not} 't Hooft coupling). 
In terms of the intensive variables the equation of state can be
written as,
\begin{equation}\label{eqstate}
P = \frac{\rho-5\rho_0}{4}, \quad s = 5 \lambda^{\frac{1}{5}}\left(\frac{\rho-\rho_0}{4}\right)^{\frac{4}{5}},
\end{equation}
where $P$, $s$ and $\rho$ are respectively the pressure, entropy density and energy density.

We now consider the energy momentum tensor as worked out in 
\cite{Lahiri:2007ae}. In a frame where the four velocity is $u^{\mu}=\gamma(1,0,\omega,0)$, the perfect 
fluid stress tensor and the surface stress tensor are respectively given by,
\begin{equation}
T^{\mu\nu}_{\text{perfect}}=\left(
	\begin{array}{llll}
	 \gamma ^2 \left(P r^2 \omega^2+\rho \right) & 0 & \omega \gamma ^2
	   (P+\rho ) & 0 \\
	 0 & P & 0 & 0 \\
	\omega \gamma ^2 (P+\rho ) & 0 & \frac{\gamma ^2 \left(P r^2
	  \omega^2+\rho \right)}{r^2} & 0 \\
	0 & 0 & 0 & P
	\end{array}
	\right)
\end{equation}

\begin{equation}
T^{\mu\nu}_{\text{surface}}=\frac{\delta(z-h(r)) \sigma  }{\sqrt{h'(r)^2+1}}
				\left(
				\begin{array}{llll}
			 	h'(r)^2+1 & 0 & 0 & 0 \\
			 	0 & -1 & 0 & -h'(r) \\
			 	0 & 0 & -\frac{h'(r)^2+1}{r^2} & 0 \\
			 	0 & -h'(r) & 0 & -h(r)^2
				\end{array}
				\right)
\end{equation}
where $\gamma=(1-\omega^2r^2)^{-\frac{1}{2}}$, P is the local pressure of the fluid configuration
and $\sigma$ is the surface tension .
The hydrodynamic equations  given by $\nabla_{\mu}T^{\mu\nu}=0$ reduces to,

\begin{equation}\label{difeqomega}
\frac{\partial P}{\partial r}-\frac{\omega^2r(\rho+P)}{\sqrt{1 - \omega^2r^2}}�\mp 2\sigma H h'(r)\delta(z-h(r))=0,
\end{equation}

\begin{equation}\label{difeqpressure}
\frac{\partial P}{\partial z}\pm 2\sigma H \delta(z-h(r))=0.
\end{equation}
where,
\begin{equation}\label{theH}
H= \mp \frac{h'(r) \left(h'(r)^2+1\right)+r h''(r)}{2 r
   \left(h'(r)^2+1\right)^{3/2}}
\end{equation}
the upper sign referring to the upper($z > 0$) surface.

We do not consider the dissipative terms in the stress tensor because 
we are interested in non-dissipative solutions. The dissipative part 
of the stress tensor vanishes on the solutions constructed in 
\cite{Lahiri:2007ae} (those that are considered by us in this paper).

We define certain dimensionless variables as follows,
\begin{equation}
\tilde{\omega}=\frac{\sigma \omega}{\rho_0},\quad v=\omega r,\quad\tilde{h}(v)=\omega h(r),
\end{equation}
where $\omega$ is the angular velocity of the fluid configuration.

Now integrating \eqref{difeqomega} and \eqref{difeqpressure} in the bulk interior away from the 
outer or inner surface (and also using the equation of state \eqref{eqstate}) we have,
\begin{equation}\label{bulk}
 P = \rho_0 \left(\frac{k}{\left(1-v^2\right)^{\frac{5}{2}}} - 1 \right)
\end{equation}
where $k$ is an integration constant depending on the bulk properties of the plasma fluid.

Again integrating \eqref{difeqpressure} across an outer surface gives,
\begin{equation}\label{outer}
P=2\sigma H
\end{equation}

For convenience we define two functions,
\begin{equation}
f(v)=2k-3(v^2+2c)(1-v^2)^{\frac{3}{2}}, \quad g(v)=6\tilde{\omega}v(1-v^2)^{\frac{3}{2}}.
\end{equation}

Now using \eqref{bulk} and \eqref{theH} into \eqref{outer} we obtain a 
second order differential equation for the height function $\tilde{h}(v)$.
This equation can be integrated once to obtain,
\begin{equation}
\label{hdash}
\frac{d\tilde{h}}{dv}=\frac{-f(v)}{(g(v)^2-f(v)^2)^{\frac{1}{2}}},
\end{equation}
which on further integration yields,
\begin{equation}
\label{hfunc}
\tilde{h}(v)=\int_{v_o}^{v}\frac{-f(x)}{(g(x)^2-f(x)^2)^{\frac{1}{2}}} dx.
\end{equation}
Here $c$ is another constant determined
by the surface properties of the fluid.

If there is an inner surface of the plasma configuration then it is 
also obtained by the same procedure, but this time we 
have the opposite sign, 
\begin{equation}\label{inner}
P=-2\sigma H
\end{equation}
Again integrating \eqref{inner} we obtain the height function of an inner surface,
which is given by,
\begin{equation}
\label{hdashI}
\frac{d\tilde{h}}{dv}=\frac{f(v)}{(g(v)^2-f(v)^2)^{\frac{1}{2}}}.
\end{equation}

Now let us consider the various stationary solutions presented in \cite{Lahiri:2007ae}
one by one.

\begin{enumerate}
\item $\mathbf{BALL}$ : This is obtained when the derivative of the height function is 0 at $v=0$ and $-\infty$ at the 
outer radius $v=v_o$ and the height function is monotonic in this interval of interest. The first condition implies
$k=3c$, the second implies $f(v_o)=g(v_o)$ and the third implies $k$ must always be greater than $1$. We also 
compute the expressions for energy, angular momentum and entropy. Energy (obtained from the time-time component of the energy momentum tensor) is given by,
\begin{equation}\label{balleng}
E = \frac{4\pi \sigma^3}{\rho_0^2 \tilde{\omega}}
    \int_{0}^{v_o}dv\left(\tilde{h}(v)v\left(\frac{k(4+v^2)}{(1-v^2)^{\frac{7}{2}}} +1\right)+
    \tilde{\omega}v(1+\tilde{h}^{\prime}(v)^2)^{\frac{1}{2}}\right).
\end{equation}

Angular momentum is obtained from the $t\phi$ component of the energy momentum tensor and is given by,
\begin{equation}\label{ballangmom}
L = \frac{20\pi k \sigma^4}{\rho_0^3 \tilde{\omega}^4}
    \int_{0}^{v_o}\frac{v^3\tilde{h}(v)dv}{(1-v^2)^{\frac{7}{2}}}.
\end{equation}

Finally the Entropy is obtained from the expression of entropy density as in \eqref{eqstate} and is given by,
\begin{equation}\label{ballent}
S = \frac{20\pi \lambda^{\frac{1}{5}}k^{\frac{4}{5}}\sigma^3}{\rho_0^{\frac{11}{5}}\tilde{\omega}^3}
     \int_{0}^{v_o}\frac{v\tilde{h}(v)dv}{(1-v^2)^{\frac{5}{2}}}.
\end{equation}

\item $\mathbf{PINCHED ~BALL}$ : This solution is topologically same as the ball. So the conditions on the derivative
of the height function is same as in the case of the ball. However the important difference here is that the
height function is not monotonic. This allows for values of k less than 1 only. The reality condition for 
the height function demands k to be greater than 0. Further restriction on the parameter space is provided by the 
fact that the height function at $v=0$ cannot exceed that at $v=v_o$ (where it is 0). In other words 
$\tilde{h}(0)$ must be greater than $\tilde{h}(v_o)=0$.
This condition may be stated as follows,
\begin{equation}\label{pballcond}
\Delta \tilde{h} = \int_{0}^{v_0}\frac{-f(x)dx}{(g(v)^2-f(v)^2)^{\frac{1}{2}}} \geq 0.
\end{equation}

The expressions for energy, angular momentum and entropy is exactly the same as in the case of the ball solution, the only difference being that here we are dealing with a different region of parameter space.

\item $\mathbf{RING}$ : In the case of the ring $\tilde{h}^{\prime}(v)$ must be $\infty$ and $-\infty$ respectively
at the inner and outer radius $v_i$ and $v_o$ (rigorously these are the velocities at the inner and outer 
radius of the ring). This sets the following conditions,
\begin{equation}\label{rcond}
\begin{split}
f(v_o) &= g(v_o),\\
f(v_i) &= -g(v_i).
\end{split}
\end{equation}
Further the condition that  $\tilde{h}(v)$ at both ends $v_i$ and $v_o$ should be equal (both being equal to 0),
sets the following condition,
\begin{equation}\label{rcondtwo}
\Delta \tilde{h} = \int_{v_i}^{v_0}\frac{-f(v)dv}{(g(v)^2-f(v)^2)^{\frac{1}{2}}} = 0.
\end{equation}
In the expressions for energy, angular momentum and entropy the lower limit of the integral gets modified for the ring
and we have,
\begin{equation}\label{elsring}
\begin{split}
 E &= \frac{4\pi \sigma^3}{\rho_0^2 \tilde{\omega}}
    \int_{v_i}^{v_o}dv\left(\tilde{h}(v)v\left(\frac{k(4+v^2)}{(1-v^2)^{\frac{7}{2}}} +1\right)+
    \tilde{\omega}v(1+\tilde{h}^{\prime}(v)^2)^{\frac{1}{2}}\right),\\
 L &= \frac{20\pi k \sigma^4}{\rho_0^3 \tilde{\omega}^4}
    \int_{v_i}^{v_o}\frac{v^3\tilde{h}(v)dv}{(1-v^2)^{\frac{7}{2}}},\\
 S &= \frac{20\pi \lambda^{\frac{1}{5}}k^{\frac{4}{5}}\sigma^3}{\rho_0^{\frac{11}{5}}\tilde{\omega}^3}
     \int_{v_i}^{v_o}\frac{v\tilde{h}(v)dv}{(1-v^2)^{\frac{5}{2}}}.
\end{split}
\end{equation}

\end{enumerate}

For convenience we define the following dimensionless quantities,
\begin{equation}
\tilde{E} = \frac{\rho_0^2 E}{4\pi \sigma^3}, \quad \tilde{L} = \frac{\rho_0^3 L}{20\pi  \sigma^4}, 
\quad \tilde{S} = \frac{\rho_0^{\frac{11}{5}} S}{20\pi \lambda^{\frac{1}{5}}\sigma^3}.
\end{equation}

We shall use these quantities for our later analysis.

\section{Existence of other possible configurations} \label{numscan}

Besides the ball, pinched ball and the ring solutions discussed in \S \ref{rev}
we can have the following topologies for the  plasma configuration

\begin{itemize}
\item Hollow ball - a ball  with a ball scooped out from inside it.
\item Hollow ring - a ring with a ring scooped out from within.
\item Torridally hollow ball - a ball with a ring cut out from inside.
\end{itemize}

The existence of the Hollow ball was already ruled out in \cite{Lahiri:2007ae}.
Thefore we consider the possibility of the existence of ring-like inner surface 
of a configuration.

If a ring is to exist as an inner surface then instead of \eqref{rcond},
it should satisfy the following condition,
\begin{equation}\label{ricond}
\begin{split}
f(v_o) &= -g(v_o),\\
f(v_i) &= g(v_i).
\end{split}
\end{equation}

The condition \eqref{rcondtwo} remains unchanged.
We considered $k$, $v_i$ and $v_o$ as the parameters of the ring.
Actually there should be two parameters. However the relation \eqref{rcondtwo}
could not be easily used to express $k$ in terms of the other parameters.
Then we numerically scanned the three dimensional parameter space 
from 0 to 1 in all directions (note that although 1 is not the upper limit of $k$,
for the other two parameters it is). All the parameters were varied by an 
interval of $0.01$. With these values of the parameters we evaluated $\Delta \tilde{h}$.
We found (for $v_i \neq v_o$) $\Delta \tilde{h}$ was negative (except when 
$v_i$ was very close to zero and $v_o$ was very close to one-when the numerical 
computations were no longer reliable).
This eliminates (up to numerical accuracies of the scan) the possibility of the 
existence of a ring like inner surface both inside a ball as well as a ring.

\section{Analytical Results} \label{analyt}

In this section we study the solutions presented in \cite{Lahiri:2007ae} in certain analytically tractable limits.

\subsection{The Thin Ring Limit}\label{thinring}

It was possible to obtain an analytical solution of the hydrodynamic equations in the thin ring limit (i.e. $(v_o-v_i)$
is very small) as we shall discuss in this subsection. Let us define the quantities,

\begin{equation}
a = \frac{v_o+v_i}{2} ; \quad \alpha a = \frac{v_o-v_i}{2}.
\end{equation}

It is sometimes convenient to choose $a, \alpha$ as the two parameters in the problem. The thin ring limit 
then corresponds to choosing $\alpha$ to be an infinitesimal parameter.
In terms of $a, \alpha$ the equations \eqref{rcond} and \eqref{rcondtwo} that constrained the parameter space of the 
ring solution may be written as,

\begin{equation}
\label{consta}
f(a(1 + \alpha)) = g(a(1 + \alpha)),
\end{equation}

\begin{equation}
\label{constb}
f(a(1 - \alpha)) = -g(a(1 - \alpha)),
\end{equation}

\begin{equation}
\label{constc}
\int_{a(1+\alpha)}^{a(1-\alpha)} \frac{f(v) dv}{(g(v)^2-f(v)^2)^{\frac{1}{2}}} = 0.
\end{equation}
These three equations may be used to obtain $k, \tilde{\omega}, c$ in terms of $a, \alpha$.

Now, in order to satisfy these equations, let us start with the ansatz :
\begin{equation}
\label{f-g-condition}
f(a+x) = \frac{x}{a\alpha} g(a+x)
\end{equation}
This form of $f(v)$, clearly satisfies the above conditions \eqref{consta}, \eqref{constb}, \eqref{constc}, and leads to the conditions:
\begin{eqnarray}
f(a) &=& 0 \\
a\alpha f'(a) &=& g(a) \\
\frac{1}{2}(a\alpha) f''(a) &=& g'(a)
\end{eqnarray}
Similar conditions on the higher derivatives cannot be imposed since we have only 3 free parameters.
To leading order in $\alpha$ these 3 parameters are determined to be,
\begin{equation}\label{approxw}
\begin{split}
k &= \frac{(1-a^2)^{\frac{7}{2}}}{1-6a^2} + O(\alpha),\\
\tilde{\omega} &= \frac{5a^3\alpha}{1-6a^2} + O(\alpha^2),\\
c &= \frac{20a^4 - 7a^2 +2}{6(1 - 6a^2)} + O(\alpha).
\end{split}
\end{equation}

To estimate the errors in this calculation we must note that the first term in $f(a+x)$ that does not match correctly is the term $\frac{1}{2}g^{(2)}(a)\frac{x^3}{a\alpha}$.  This term does not contribute to any change in the numerator of the integral, however, in the denominator it contributes a deviation of $O\left(x^3/\alpha \right)$,
which lends to an error of $O\left(\alpha^2 \right)$ overall.

Also under the small $\alpha a$ approximation the height function can be approximated as,
\begin{equation}
\label{aproxh}
\begin{split}
\tilde{h}(v) = \tilde{h}(x+a) =& \int_{-\alpha a}^{x} \frac{-f'(a)xdx}{\big(g(a)^2-f'(a)^2x^2 \big)^{\frac{1}{2}}}\\
              =& \big(\alpha^2 a^2 - (v-a)^2 \big)^{\frac{1}{2}},
\end{split}
\end{equation}
where we use $g(a)=(\alpha a)f'(a)$.

Now the angular momentum $(\tilde{L})$ energy $(\tilde{E})$ and entropy $(\tilde{S})$ in the small $\alpha$ limit is given by,
\begin{equation}\label{angmom}
\tilde{L} = \frac{k}{\tilde{\omega}^4} \int_{v_i}^{v_o} \frac{v^3 h(v) dv}{(1-v^2)^{\frac{7}{2}}}
  \approx \frac{\sqrt{a^2} \left(1-6 a^2\right)^3 \pi }{1250 a^8 \alpha^2}
\end{equation}

\begin{equation}\label{entropy}
\tilde{S} = \frac{k^{\frac{4}{5}}}{\tilde{\omega}^3} \int_{v_i}^{v_o} \frac{v h(v) dv}{(1-v^2)^{\frac{5}{2}}}
  \approx -\frac{\sqrt{a^2} \left(1-6 a^2\right)^2 \left(a^2-1\right) \pi
   }{250 a^7 \sqrt[5]{-\frac{\left(1-a^2\right)^{7/2}}{6 a^2-1}}
   \alpha (a)}
\end{equation}

\begin{equation}\label{energy}
\begin{split}
\tilde{E} &= \frac{1}{\tilde{\omega}^3}\int_{v_i}^{v_o} dv \left( v h(v)\left(\frac{(4+v^2)k}{(1-v^2)^{\frac{7}{2}}}
 + 1 \right) + \frac{\tilde{\omega} v a \alpha}{h(v)} \right)\\
  &\approx \frac{\sqrt{a^2} \left(1-6 a^2\right)^2 \left(a^2+1\right) \pi
   }{50 a^7 \alpha}
\end{split}
\end{equation}

The local temperature of the plasma is given by,
\begin{equation}
\mathcal{T} = \left(\frac{\rho-\rho_0}{4\lambda}\right)^{\frac{1}{5}} 
= \left(\frac{k\rho_0}{\lambda}\right)^{\frac{1}{5}}\gamma,
\end{equation}
where $\gamma=\frac{1}{\sqrt(1-v^2)}$, $v$ being the local velocity of the fluid.

Also the thermodynamic temperature of full plasma configuration is given by,
\begin{equation}
T = \left(\frac{\partial E}{\partial S}\right)_{L} 
  = \left(\frac{\partial \tilde{E}}{\partial \tilde{S}}\right)_{\tilde{L}}\frac{1}{5}
\left(\frac{\rho_0}{\lambda}\right)^{\frac{1}{5}}.
\end{equation}

Now using relations \eqref{angmom}, \eqref{entropy} and \eqref{energy} we obtain the temperature to leading 
order in $\alpha$ as follows,
\begin{equation}
\frac{1}{T}=\beta=\left(\frac{\delta \tilde{S}}{\delta \tilde{E}}\right)_{\tilde{L}} = \frac{1}{5k^{\frac{1}{5}}}
\end{equation}
so that temperature of the configuration is given by,
\begin{equation}
T = \mathcal{T}\sqrt{1-v^2}.
\end{equation}

We can now use \eqref{energy} to express $\alpha$ in terms of $a$ and energy $E$, 

\begin{equation}
\label{alpha}
\alpha = \frac{\sqrt{a^2} \left(1-6 a^2\right)^2 \left(a^2+1\right) \pi
   }{50 a^7 \tilde{E}}
\end{equation}

\begin{figure}[htp]
\centering
\includegraphics[width=0.4\textwidth]{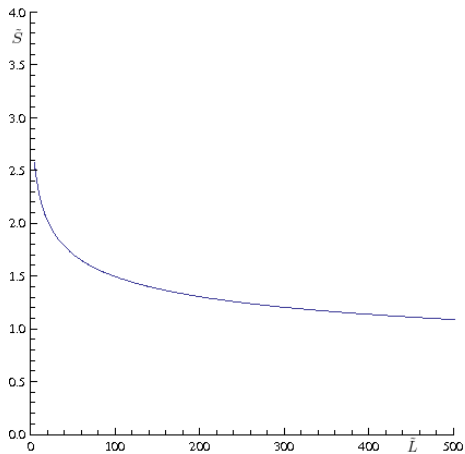}
\caption{ Here we plot entropy ($\tilde{S}$) against angular momentum ($\tilde{L}$) for $\tilde{E}=10$ 
 (note that even when the thin ring approximation is invalid for such low energies this plot agrees quite 
 well with the numerical plots presented in \S5.}
\label{fig:lvsse10}
\end{figure}

\begin{figure}[htp]
\centering
\includegraphics[width=1\textwidth]{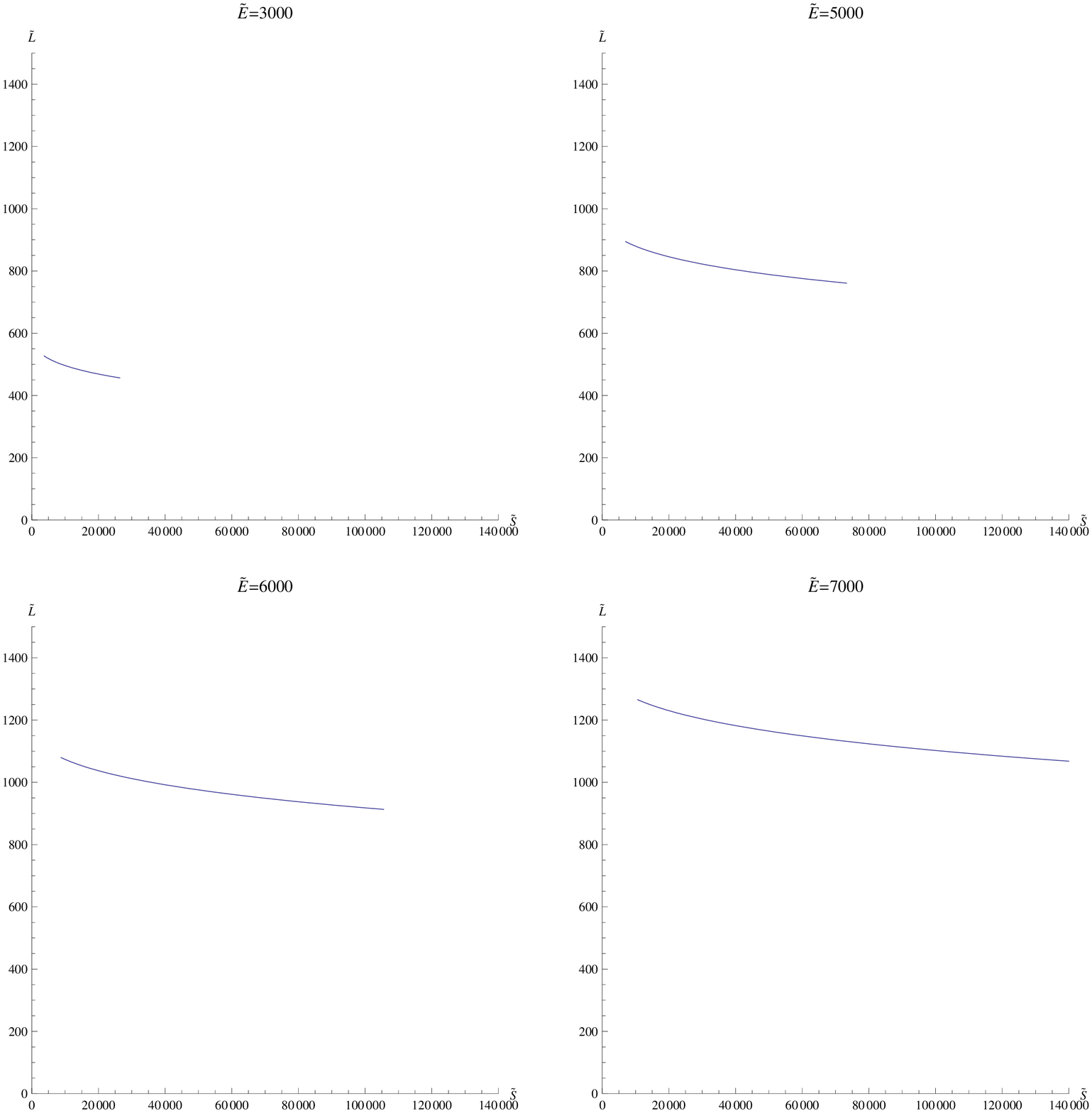}
\caption{Here we again plot entropy ($\tilde{S}$) against angular momentum ($\tilde{L}$) for 
different fixed energies ($\tilde{E} = 3000, 5000, 6000, 7000$) within the domain of validity 
of the thin ring approximation. for all the plots  the left end is the point where 
$\alpha$ is 0.1 and the right end is the point where the hydrodynamics 
assumption starts breaking down.
}
\label{fig:lvssmanye}
\end{figure}

Substituting $\alpha$ from \eqref{alpha} into \eqref{angmom} and \eqref{entropy} we are able 
to express $\tilde{L}$ and $\tilde{S}$ in terms of $\tilde{E}$ and $a$. Thus we can now fix $\tilde{E}$ 
and plot $\tilde{S}$ vs $\tilde{L}$ taking 
$a$ to be a parameter. This plot is shown in fig:\ref{fig:lvsse10} and in fig:\ref{fig:lvssmanye} for 
different values of energy.

\subsubsection{Validity of the Thin Ring Approximation}

Our conclusions about the thin ring is valid only when we simultaneously 
meet the following conditions.
\begin{enumerate}
\item the thin ring limit is valid  i.e. $\alpha \ll 1$
\item physical quantities like $\tilde{\omega}$ and $\tilde{L}$ are not negative.
\item we are within the domain of validity of hydrodynamics $\frac{\sigma}{\rho_0 R_c} \ll 1$.
This physically means that the radius of the thin ring considered ($R_c$) is much larger than
$\frac{\sigma}{\rho_0}$, the typical length scale at  which the hydrodynamics 
assumption breaks down.
\end{enumerate}
We now analyze the circumstances under which all of these conditions are simultaneously 
fulfilled. In the following discussion we parameterize our ring solution by its mean velocity $a$ and its energy ${\tilde E}$. 

The second condition listed above requires that $a \leq \frac{1}{\sqrt{6}}$ (see \eqref{approxw}
\eqref{angmom} ). The third condition - the validity of the fluid dynamical approximation -
imposes the more stringent requirement $a\ll\frac{1}{\sqrt{11}}$. Finally the 
condition that $\alpha \ll 1$ is met provided 
$$ \frac{(1-6a^2)^2}{a^6 {\tilde E}} \ll 1 .$$ 
It follows that our analysis is valid in every respect only if $a \ll 1$ but 
$a^6 {\tilde E} \gg 1$.

We conclude that the thin ring approximation described in this section is valid over a large region of parameter space provided that the energy ${\tilde E}$ is large. We
illustrate this point in fig:\ref{fig:lvssmanye}. In fig:\ref{fig:lvssmanye} we plot the entropy versus the angular 
momentum of the ring solutions at different values of ${\tilde E}$ using the thin ring formulas 
\eqref{angmom}, \eqref{entropy}, \eqref{energy}, and indicate the approximate 
region of validity of this approximation. 
As is apparent from this figure, the domain of validity of the thin ring approximation 
increases with increasing energy.

We note an intriguing property of the equations in the thin ring limit; they suggest 
a universal upper bound of $\frac{1}{\sqrt{6}}$ for the mean ring velocity. As $a$ is taken to this upper bound value the thickness of the ring $\alpha a$ goes to zero. Unfortunately- as we have seen above - this phenomenon occurs outside the validity of the fluid dynamical approximation and so is not reliable. Nonetheless it is natural to 
wonder whether an exact version of this phenomenon is present for the true
solutions of gravity.

\subsubsection{Physical Interpretation}

As we have seen in the previous subsection, the conclusions that follow from an analysis 
of the thin ring limit are valid - within the regime of fluid dynamics - only at small 
$a$. In this regime the fluid is approximately non relativistic, and all the formulas 
of the previous subsection may be derived in a much more elementary and physical fashion, as we 
will now explain. 

In the thin ring limit we can think of the solution as a spinning narrow tube in the shape of a torus, parameterized by the two radius $R_c$ (radius of the tube) and $R$ (distance of the center of the tube from the center of the torus). We are now interested in the limit $\frac{R_c}{R} \rightarrow 0$. In this limit the plasmaring is a narrow 
tube that curves into a circle that is much larger than its thickness. As a consequence it is possible to focus 
on a segment of this ring that is large compared to $R_c$ but small compared to $R$. To the zeroth approximation
such a segment may be regarded as a cylinder of radius $R_c$. A simple force balance (surface tension versus 
pressure) on this cylinder tells us that the pressure of the fluid in this cylinder is given by
\begin{equation}
P = \frac{\sigma}{R_c}
\end{equation}

Now the cylinder described above is not quite straight; it bends in the plane of the ring and also rotates 
in the same plane. Consequently, the condition for equilibrium at next order in an expansion of $R_c/R$ require us 
to balance forces in the plane of the ring. This is also easily achieved. The centripetal acceleration of rotation, on any
given segment of the ring, must be provided by the effective tension of the cylinder. This tension is given by 
$ T=2 \pi R_c \sigma  -\pi R_c^2 P= \pi R_c \sigma$. A force balance then yields $T= \pi R_c^2 R^2 \rho \omega^2$ so that  
\begin{equation}
\omega^2=\frac{ \sigma}{\rho R_c R^2},
\end{equation}
Expressing this relation in terms of the quantities defined previously we obtain
\begin{equation}
\tilde{\omega} = \frac{5a^3\alpha}{1-4a^2} \approx 5 a^3 \alpha,
\end{equation}
which agrees with \eqref{approxw} at leading order in $a$ (we could not have expected our 
better agreement from our nonrelativistic arguments; see below however for the relativistic 
generalization). Now we are interested in the behavior of angular momentum $L$ and entropy $S$ keeping energy $E$ fixed in the limit $R_c \rightarrow 0$. The energy expressed in terms of $a$ and $\alpha$ up to leading order is (we do not include the surface contribution because it is not expected to contribute to the leading order),
\begin{equation}
\label{eng}
E = \rho (2 \pi R_c^2 R) + \frac{1}{2} \rho (2 \pi R_c^2 R) ( \omega^2 R^2 ) \approx \frac{\pi}{50 a^6 \alpha},
\end{equation}
which again agrees to our result in \eqref{energy} up to leading order in $a$. 

Similar agreements are obtained for angular momentum and entropy  which are obtained to be respectively,
\begin{equation}
L= \int \rho dv \omega R^2 \approx \frac{\pi}{1250 a^7 \alpha^2} ,
\end{equation}
and 
\begin{equation}
S = \frac{5 \lambda^{1/5} (\rho - \rho_0)^{4/5} V}{4^{4/5}} \approx \frac{\pi}{250 a^6 \alpha}.
\end{equation}

Although we are never really able to go beyond small $a$ within the validity of fluid 
dynamics, it is of course true as a purely mathematical fact that a relativistic 
version of the force balance presented in this subsection may be used to reproduce 
the exact formulas (to all orders in $a$) of the thin ring approximation. In the rest 
of this subsection we describe how that goes.

Now for our case of the thin ring the height function is given by (as in \eqref{aproxh}).
\begin{equation}
h(r)=\sqrt{R_c^2-(r-R)^2}.
\end{equation}

Using this expression for $h(r)$ into \eqref{theH} and \eqref{outer}
we obtain the pressure to be $P=\frac{\sigma(2r-R)}{2rR_c}$. We now plug this back into \eqref{difeqomega} and
consider the point $z=0,\ r=R$. Then if we write the resultant equation in terms of the previously defined parameters 
$\alpha,a$ we obtain,
\begin{equation}
\tilde{\omega}=\frac{5a^3\alpha}{1-6a^2},
\end{equation}
which agrees completely to \eqref{approxw} up to the leading order in $\alpha$.
Further using the equation of state we can integrate \eqref{difeqomega} to obtain,
\begin{equation}
k=\frac{(\rho-\rho_0)(1-v^2)^{\frac{5}{2}}}{4\rho_0},
\end{equation}
$k$ here being the integration constant. We can once again use the equation of state and the expression of pressure
obtained above in to this expression for k. Then if we express it in terms of $\alpha,a$ we obtain an exact match with
\eqref{approxw} up to leading order in $\alpha$. Once the expressions for $\tilde{\omega}$ and $k$ match 
up to leading order in $\alpha$ the matching of the expressions for energy, entropy and angular momentum follow
immediately.

\subsubsection{Thin Ring as an Inner Surface}

We now investigate whether a thin ring solution may exist as a inner surface inside a outer surface being a ball 
or a ring. If the thin ring is to exist as a inner surface the conditions corresponding to \eqref{consta} 
and \eqref{constb} is changed to,
\begin{equation}\label{condI}
f(a(1-\alpha))=g(a(1-\alpha)),
\end{equation}
and,
\begin{equation}\label{condII}
f(a(1+\alpha))=-g(a(1+\alpha)),
\end{equation}
whereas the condition \eqref{constc} remains unchanged (note the all important change in sign). 

Note that the constant c that appears in the above equations is 
in general different from the similar constant that will appear for the outer surface. However Since in our argument 
bellow we never refer to the outer surface so we continue to use the notation c for this integration constant
determined by the surface properties. 

As before we can expand \eqref{condI} and \eqref{condII} in terms of the parameter $\alpha$ which is small by 
the thin ring assumption. Then we get,
\begin{equation}
g'(a)=-\frac{1}{2}f''(a)\alpha a,
\end{equation}
\begin{equation}
g(a)+\frac{1}{2}g''(a)(\alpha a)^2 = -f'(a)\alpha a. 
\end{equation}
Also as before we have $f(a)=0$.
Solving the above three conditions for $k,\tilde{\omega}$ and c we get the following result,
\begin{equation}
\begin{split}
k &= \frac{\left(1-a^2\right)^{5/2} \left(\left(84 \alpha ^2-2\right)
   a^6+\left(6-75 \alpha ^2\right) a^4+\left(9 \alpha
   ^2-6\right) a^2+2\right)}{12 \left(2 \alpha ^2-1\right)
   a^6+\left(26-30 \alpha ^2\right) a^4+\left(9 \alpha
   ^2-16\right) a^2+2} \\
\tilde{\omega}&= -\frac{10 a^3 \left(a^2-1\right)^2 \alpha }{12 \left(2 \alpha
   ^2-1\right) a^6+\left(26-30 \alpha ^2\right) a^4+\left(9
   \alpha ^2-16\right) a^2+2}\\
c&= \frac{-40 \left(6 \alpha ^2-1\right) a^8+\left(408 \alpha
   ^2-94\right) a^6+\left(72-195 \alpha ^2\right) a^4+2 \left(9
   \alpha ^2-11\right) a^2+4}{6 \left(12 \left(2 \alpha
   ^2-1\right) a^6+\left(26-30 \alpha ^2\right) a^4+\left(9
   \alpha ^2-16\right) a^2+2\right)} .
\end{split}
\end{equation}
We expand the expressions of $k$ and $\tilde{\omega}$ up to leading order in $\alpha$ to obtain,
\begin{equation}
\begin{split}
k&=\frac{\left(1-a^2\right)^{7/2}}{1-6 a^2} \\
\tilde{\omega}&=\frac{5 a^3 \alpha }{6 a^2-1} .
\end{split}
\end{equation}
Thus we see that if $a>\frac{1}{\sqrt{6}}$, $\tilde{\omega}$ is positive but $k$ is negative; on the other hand if 
$a<\frac{1}{\sqrt{6}}$ (especially if $a<\frac{1}{\sqrt{11}}$ where our hydrodynamic approximation is valid)
$k$ is positive but $\tilde{\omega}$ is negative. Thus for a given $\alpha$ (or energy) there is no velocity $a$
for which a solution with a thin ring as a inner surface exists.

\subsection{The Pinched Limit Of The Ring}

As we have explained above, the fluid dynamical equations under study in this paper admit two topologically 
distinct classes of solutions; solid balls and solid tori. The moduli space of ball and ring solutions 
meet at a point which we call the pinched limit. In this limit the inner radius of the ring goes to zero and 
the ring closes off into a ball. Approached from the other side, the height function of the ball undergoes 
an extreme pinch (goes to zero at the center) at this point, opening out into a ring. In this section we study 
the ring solution in the neighborhood of this topological transition.

\begin{figure}[htpb]
\begin{center}
$\begin{array}{c@{\hspace{1in}}c}
\includegraphics[width=0.55\textwidth]{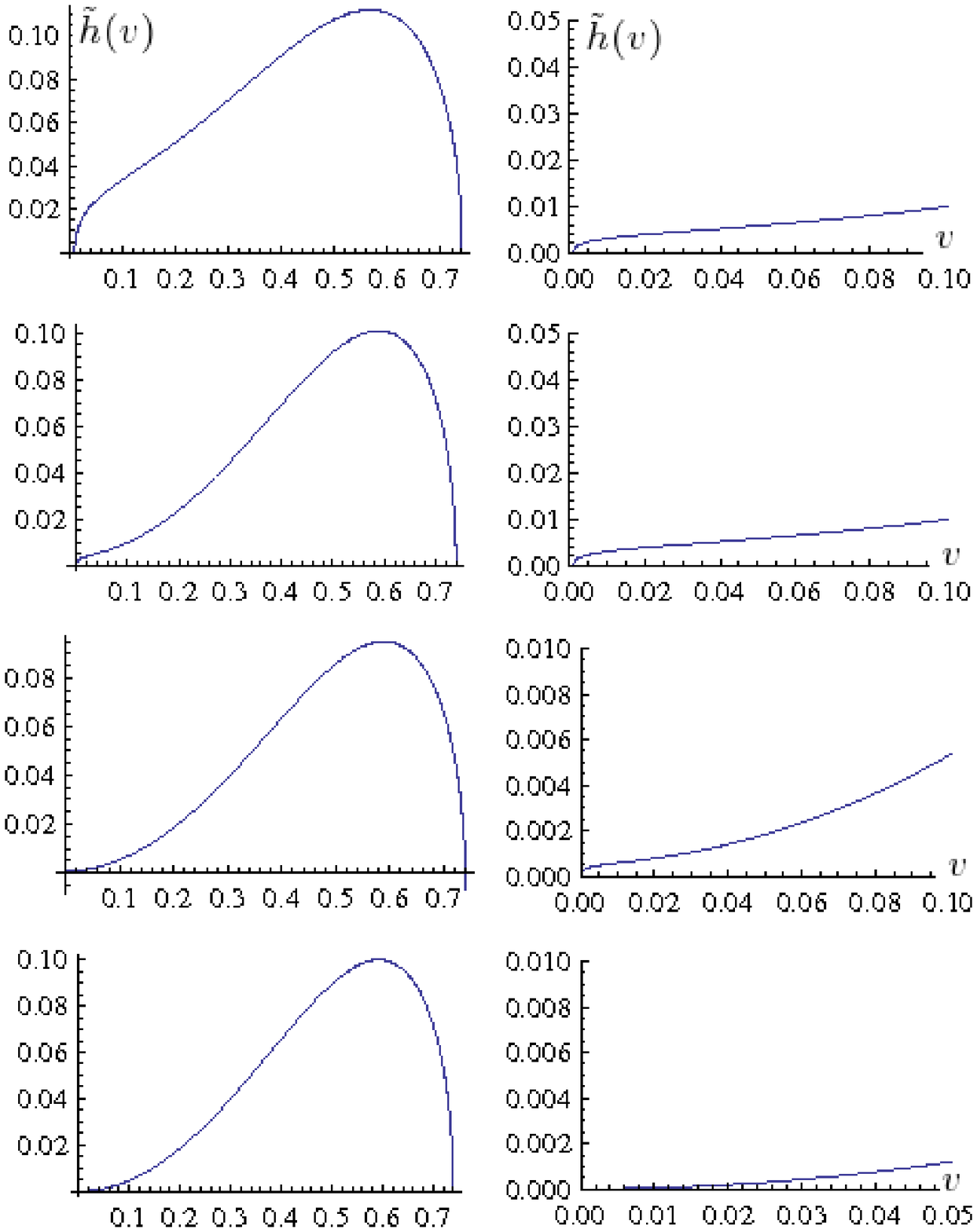} &
	\includegraphics[width=0.3\textwidth]{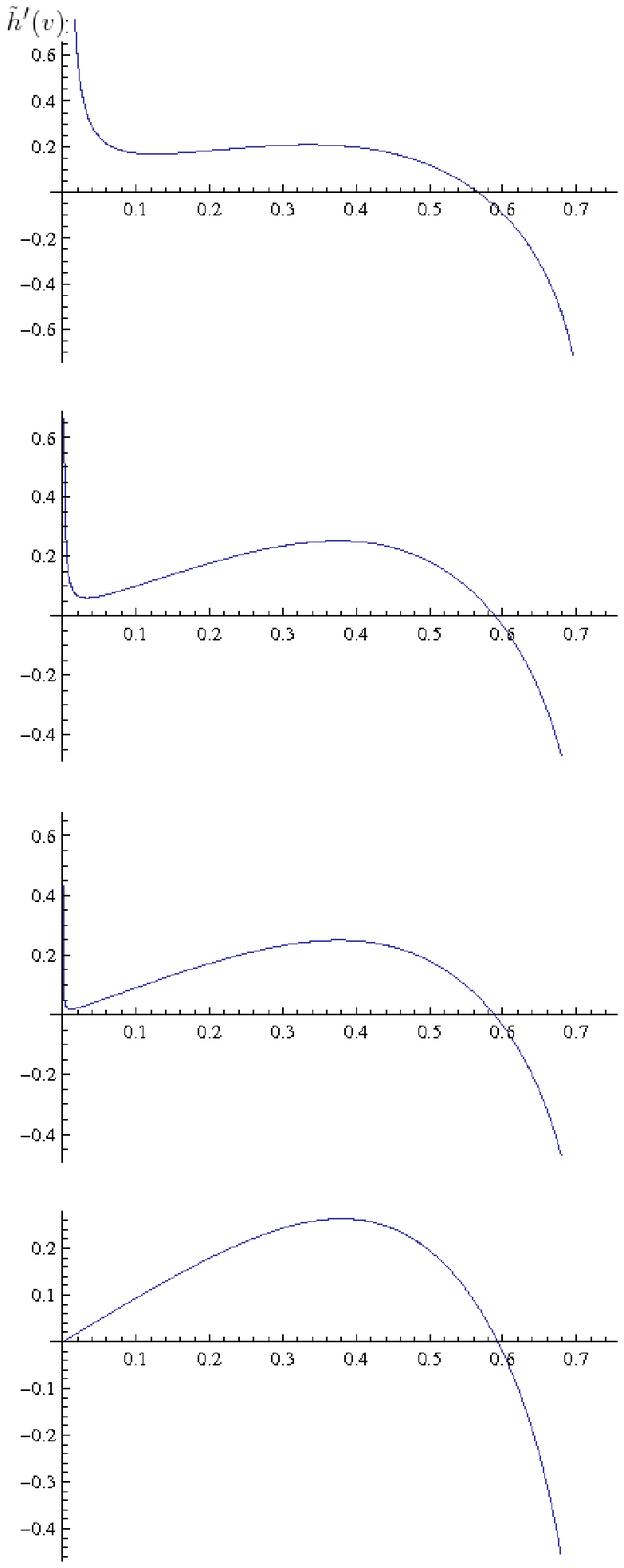} \\ [0.4cm]
\mbox{\bf (a)} & \mbox{\bf (b)}
\end{array}$
\end{center}
\caption{In $(a)$ we plot $\tilde{h}(v)$ (the plot at the side shows $\tilde{h}(v)$ near $v=v_i$) and in $(b)$
we plot $\tilde{h}^\prime(v)$. All the plots are for $v_o=0.74$ and the value of $v_i$ are
 $0.01, 0.001, 0, 0001, 10^{-14}$ as we go from top to bottom.}
\label{fig:handhdash}
\end{figure}

In the neighborhood of the pinch, i.e. when $v_i<v \ll 1$, 
\begin{equation}\label{hdashII}
\tilde{h}'(v) =
 \frac{6\tilde{\omega}v_i+6(1-k)v_i(v-v_i)+3(1-k)(v-v_i)^2}{\sqrt{36\tilde{\omega}^2(v-v_i)^2+72\tilde{\omega}
^2(v-v_i)v_i}},
\end{equation}
where we have retained terms up to second order in both $v-v_i$ and $v_i$. It is not difficult to convince  
for small $v_i$,
\begin{equation}\label{hdashIII} 
\begin{split}
\tilde{h}'(v) & = \sqrt{\frac{v_i}{2}}\frac{1}{\sqrt{v-v_i}},~~~\text{when}~(v-v_i) \leq 2 v_i.\\
\tilde{h}'(v) & = \frac{v_i}{v-v_i},~~~\text{when}~2 v_i \leq (v-v_i) \leq 
                                           \left(\frac{2\tilde{\omega} v_i}{(1-k)}\right)^\frac{1}{2}\\ 
\tilde{h}'(v) & = \frac{(1-k)}{2\tilde{\omega}}(v-v_i),~~~\text{otherwise}.
\end{split}
\end{equation}
Note in particular that when $v_i$ is set to zero  
\begin{equation}
\tilde{h}(v) = \frac{(1-k)}{4\tilde{\omega}}v^2.
\end{equation}
smoothly matching on the the ball in the limit of extreme pinching. 
In fig:\ref{fig:handhdash} we plot the height function and its derivative 
at various values of $v_i$.

The transition from the ball to the ring is continuous in configuration space - it is the analogue of a 
second order phase transition. It would not be difficult to extract the `critical exponents of this 
transition from the formulas presented in this paper. However it seems likely that the thin wall approximation
used to model the fluid boundary in this paper (see \cite{Aharony:2005bm, Lahiri:2007ae}
for more details) is inadequate precisely in the strict limit $v_i \to 0$.
At least naively we would expect the thin wall approximation to be good only when $v_i \gg \tilde{\omega}$.

\section{Numerical Results} \label{num}

\subsection{Thermodynamics of Plasma Balls}

\begin{figure}[htpb]
\centering
\includegraphics[width=0.3\textwidth]{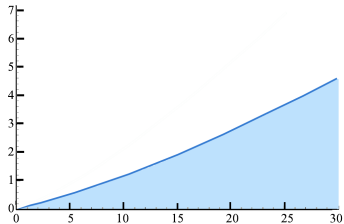}
\caption{Allowed region of energy (x-axis) and angular momentum (y-axis) space for ordinary ball solution. 
}
\label{fig:ordb}
\end{figure}

In \S \ref{rev} we had presented the various solutions along with the expression 
for their energy, angular momentum and entropy. 
For the ball-type solution we may take the two independent parameters 
characterizing the solution to be energy($\tilde{E}$) and angular momentum($\tilde{L}$). Now  
\eqref{balleng} and \eqref{ballangmom} express $\tilde{E}$ and $\tilde{L}$ as functions of two auxiliary 
parameters $k$  and $v_o$. Consequently by plotting $\tilde{E}$ and $\tilde{L}$ for any given value of 
$k$ and $v_o$, and by varying  these parameters over their allowed region, we generate the region of the $\tilde{E}$-$\tilde{L}$ space for which we have ball type solutions. 
For the ordinary ball we have $k \geq 1$ and $v_o \in [0,1]$. With these ranges for $k$ 
\footnote{we considered $k$ up to 7.5  only. With increase in the value of $k$ the region near the energy(x)-axis
(i.e. lower values of angular momentum)
gets filled; therefore the fact that an upper bound to angular momentum exists at a fixed energy can be
asserted with confidence.} and $v_o$ we plot the points generated for 
$\tilde{E}$ and $\tilde{L}$ on a diagram in fig:\ref{fig:ordb}.

\begin{figure}[htpb]
\begin{center}
$\begin{array}{c@{\hspace{1in}}c}
\includegraphics[width=0.3\textwidth]{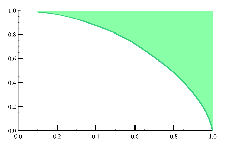} &
	\includegraphics[width=0.3\textwidth]{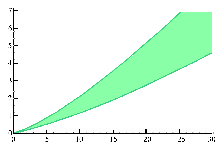} \\ [0.4cm]
\mbox{\bf (a)} & \mbox{\bf (b)}
\end{array}$
\end{center}
\caption{In $(a)$ we show the allowed region of the $v_o$ (x-axis) and k (y-axis) space where the pinched ball solution exists and in $(b)$ we show the corresponding allowed region of energy (x-axis) and angular momentum (y-axis) space for pinched ball solution.}
\label{fig:pbkvandel}
\end{figure}

For the pinched ball solution $k \in [0,1]$ and so does $v_o$. However there is also 
the additional constraint \eqref{pballcond}. Taking this condition into account we plot the 
allowed values of $k$ and $v_o$ in fig:\ref{fig:pbkvandel}(a). Then Using these allowed 
values we generate the set of allowed points in the $\tilde{E}$-$\tilde{L}$ space and present it in a 
plot in fig:\ref{fig:pbkvandel}(b).

As is clear from this figure, ball solutions exist in a certain region of the $\tilde{E}$ , $\tilde{L}$ plane, 
and at no value of energy and angular momentum do we have more than one ball solution. 

Using \eqref{ballent} we may now compute the entropy of the ball solutions as a function of their energy and 
angular momentum. In  fig:\ref{fig:ballringlsplots} (see the left inset) we present a slice of this 
plot at fixed energy $\tilde{E}=10$ (for convenience we
found a set of values of $k$ and $v_o$ with $\tilde{E}=10 \pm 0.05$ in order to generate this plot).

\subsection{Thermodynamics of Plasma Rings}

\begin{figure}[htpb]
\centering
\includegraphics[width=0.3\textwidth]{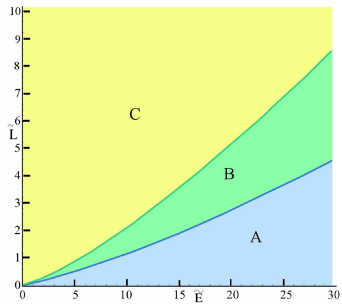}
\caption{Here we present the $\tilde{E} \ \tilde{L}$ plane showing regions where 
the various solutions viz. the ordinary ball, the pinched ball and the ring exists.
In region $A$ we have only a single ball solution.
In region $B$ we have one ball (either ordinary or pinched) and two ring solutions. 
In region $C$ we have a single (thin) ring solution.
}
\label{fig:ballringleplots}
\end{figure}

We have performed an analysis similar to the one described in the last subsection in order to obtain 
the region of the ${\tilde E} \tilde{L}$ plane in which ring solutions exist. Once again solutions exist only 
in a sub space of the ${\tilde E} \tilde{L}$ plane (for any given energy there is a minimum angular momentum
above which we have ring type solutions, see fig:\ref{fig:ballringleplots}).
There are also regions in the ${\tilde E} \tilde{L}$ plane where 
there exist two distinct ring solutions (we refer to these as thin and thick rings). This region of 
multiple solutions coincides exactly with the region of the ${\tilde E} \tilde{L}$ plane in which a ball 
solutions (ordinary or pinched) exists overlaps with that of the ring solution.

Putting all this together with the results of the the previous section we see that the ${\tilde E} \tilde{L}$ 
plane is divided into three distinct regions, $A, B, C$ as shown in fig:\ref{fig:ballringleplots}. 
In region $A$ we have only a single ball solution.
In region $B$ we have one ball and two ring solutions. In region $C$ we have a single (thin) ring solution.

\begin{figure}[htpb]
\centering
\includegraphics[width=0.7\hsize]{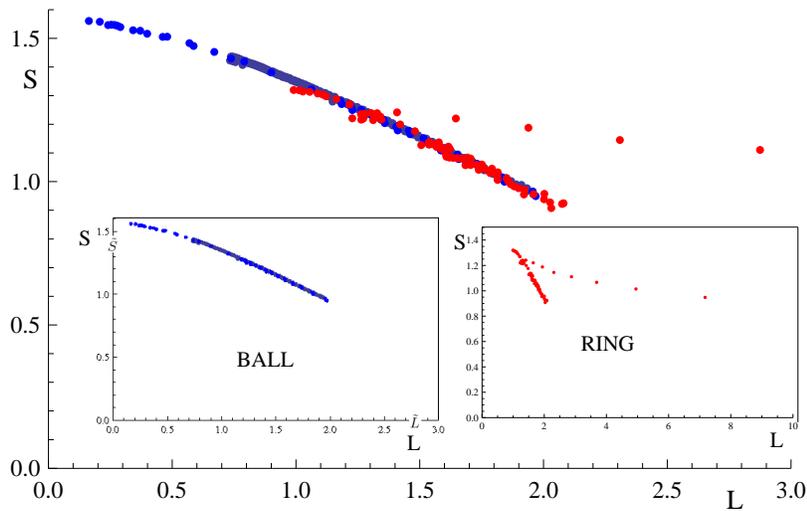}
\caption{Plot showing $\tilde{S}$(y-axis) vs $\tilde{L}$(x-axis) for all the solutions together.}
\label{fig:ballringlsplots}
\end{figure}

In the region $B$ of fig:\ref{fig:ballringleplots}, 
the solution of maximum entropy will dominate the thermodynamics of the system. In order
to see how this goes we once again fix  ${\tilde E}=10$ and plot the entropy versus angular momentum of ring 
solutions in fig:\ref{fig:ballringlsplots}(see right inset).
In  fig:\ref{fig:ballringlsplots} we also superpose this plot with that of the ball solution to obtain the entropy
versus angular momentum plot of all solutions  at constant energy. 
The shape of this final plot is schematically depicted in fig:\ref{fig:comp}(b)
We see from this plot that the thick ring solution is always entropically subdominant compared to the 
ball and the thin ring. As we have explained in the introduction, the ball dominates at smaller angular momenta, 
while the thin ring is entropically dominant at larger angular momenta.

\section{Comparison with the results in flat space}\label{compar}

Above we have used the effective fluid dynamical description of the dynamics of the deconfined phase 
of the field theory dual to gravity on $AdS_6$ compactified on a Scherk-Schwarz circle to investigate the 
structure of large rotating black holes and black rings in this gravitational background. In this section 
we qualitatively compare our results with known results and conjectures about the structure of black holes and 
black rings in flat six dimensional space. 
 
While rotating Myers Perry black holes have been analytically constructed in flat space six dimensional 
space, exact black ring solutions have not yet been constructed. Nonetheless Emparan and collaborators \cite{Emparan:2007wm, Emparan:2008eg}
have presented physically motivated conjectures for the structure of these solutions (see also \cite{Delsate:2008kw}
for a perturbative construction of a non uniform  black string in $AdS_6$).
In this section we will compare the 
moduli space of solutions obtained in this paper with that conjectured in \cite{Emparan:2007wm}. 
We will find some similarities but also other differences.

\begin{figure}[htpb]
\begin{center}
$\begin{array}{c c}
\includegraphics[width=0.5\textwidth]{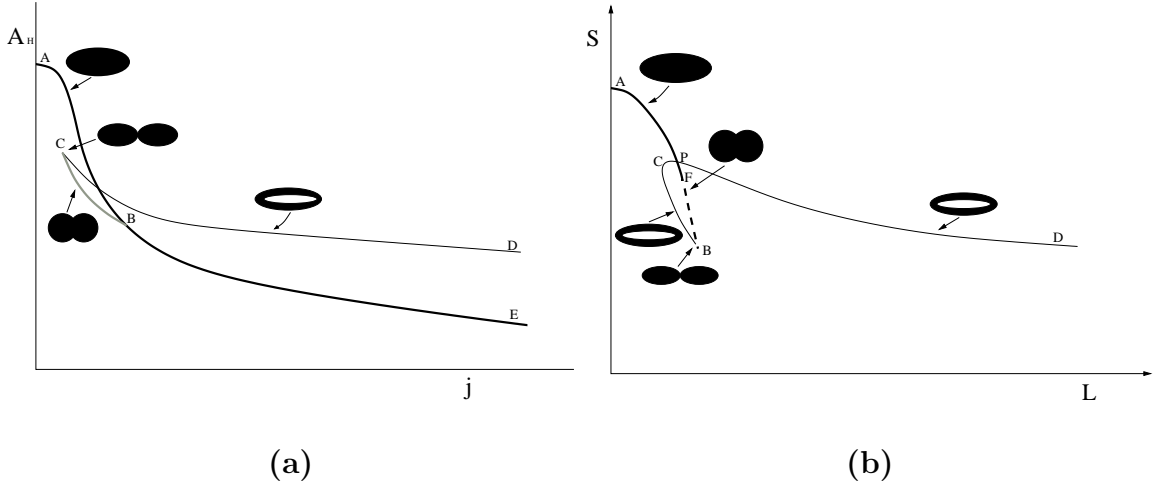} &
	\includegraphics[width=0.5\textwidth]{OurI.eps} \\ [0.4cm]
\mbox{\bf (a)} & \mbox{\bf (b)}
\end{array}$
\end{center}
\caption{In $(a)$ the area of the horizon has been plotted against angular momentum for black hole topologies 
in six dimensional asymptotically flat space as in  \cite{Emparan:2007wm} (we present only that part of the phase curve that is relevant for 
comparison with our result). In $(b)$ we summarize our result qualitatively.}
\label{fig:comp}
\end{figure}

In fig:\ref{fig:comp}(a) we present the relevant part of the conjectured phase diagram 
(the area of the horizon vs the angular momentum) in asymptotically flat 
six dimensional space. Here the dark line from A to E through B represents the well known Myers-Perry black holes 
which is a class of rotating black holes with a spherical ($S^4$-ordinary ball) horizon topology. 
The thin line from C to D
represents the black ring (with $S^3 \otimes S^1$ topology) and the grey thick line from B to C is the conjectured smooth 
interpolation from the ordinary ball to the ring type through the `pinched ball' solutions, with the  point C being the 
point of extreme pinching. It is important to note however that the authors in \cite{Emparan:2007wm} make it clear that this part of the diagram is a guess.

In fig:\ref{fig:comp}(b) we present qualitatively the phase diagram 
(Entropy vs angular momentum at constant energy) obtained by us for asymptotically $AdS_6$ spaces. Here
the dark line from A to F represents the rotating black hole with spherical topology which is the analog of 
Myers-Perry black holes in asymptotically flat space. The two phase diagrams have several points of difference. 
Firstly, at any given energy there exists a  Myers-Perry black holes at every value of angular momentum
no matter how large. In contrast, at any given energy the ball like fluid solutions determined in this paper
exist up to only a finite value of angular momentum. Also in fig:\ref{fig:comp}(b) the segment F to B 
represents the pinched ball configuration. Note that the ordinary ball smoothly continues into the 
pinched ball at F. This is unlike the flat space predictions where there is a kink at the point where the ordinary 
ball continues into a pinched ball (see point B in fig:\ref{fig:comp}(a)).

This important difference feeds into the next point of 
distinction between fig:\ref{fig:comp}(a) and fig:\ref{fig:comp}(b). 
As we have explained in our paper, the moduli space of balls ends, at 
large angular momenta, as an extreme pinched ball. This ball smoothly turns into a thick ring giving rise to 
the segment $B C$ in fig:\ref{fig:comp}(b) with point $B$ representing the 
extreme pinched ring. The natural continuation of 
fig:\ref{fig:comp}(b) to flat space would be to push the point $B$ to infinity as a consequence 
of which the Myers Perry black holes would `pinch off' only at infinite angular momentum.
This would result in a phase diagram with 
three solutions at all angular momenta larger than a critical value, 
with the thick ring always entropically subdominant, and approaching a pinch at infinite angular momenta.
In such a scenario the fact that the thin ring is entropically dominant compared to the Myers Perry black holes
is similar to that in fig:\ref{fig:comp}(a). However it would still be qualitatively different 
from fig:\ref{fig:comp}(a). This perhaps suggest that despite similarities there are considerable difference
between asymptotically AdS and flat spaces. Such differences between AdS and flat space in five dimensions 
regarding existence of Saturn type solutions have also been reported in \cite{Evslin:2008py}.
Alternatively the continuation of fig:\ref{fig:comp}(b) to flat space could be such that the ordinary ball
solution is continued from the point $F$ to infinite angular momentum such that the thin ring at large angular 
momentum is always entropically dominant compared to the ordinary ball. This diagram would be more close to 
fig:\ref{fig:comp}(a). These diagrams would have a special point where the solution with ball topology 
would split up into two lines. It would clearly be interesting to have a better understanding of this point.

\section{Discussion}

At strong coupling the AdS/CFT correspondence identifies the effectively classical large $N$ dynamics 
of an appropriate boundary theory with gravitational dynamics of the bulk dual. As field theory dynamics
reduces to fluid dynamics at long wavelengths, it follows as a prediction of the AdS/CFT correspondence 
that the appropriate gravitational equations reduce to the equations of boundary fluid dynamics in the long 
wavelength limit. For the special case that the boundary theory is conformal (and the corresponding bulk 
solutions asymptote to $AdS$ space) this equivalence has recently been proved, and the corresponding 
distinguished fluid dynamical system characterized in detail 
(see \cite{Bhattacharyya:2007vs,Bhattacharyya:2008jc,Bhattacharyya:2008xc,Bhattacharyya:2008ji}). 

In this paper we have exploited the duality between a 5 dimensional CFT on a Scherk-Schwarz circle and 
gravity on an asymptotically Scherk-Schwarz $AdS_6$ space. It follows as a consequence of the results eluded to in 
the previous paragraph, that long wavelength gravitational dynamics in this bulk background is simply dual 
to the dimensional reduction of the distinguished equations of fluid dynamics, referred to above, apart from 
a constant additive piece in the equation of state of the fluid (see \cite{Lahiri:2007ae} for details of this shift). 
This additive shift implies that the pressure of our fluid (unlike a conformal fluid) goes to zero at a particular 
finite energy density. This fact permits the existence of qualitatively new solutions of the equations 
of fluid dynamics in these systems; finite blobs of (perhaps rotating) fluids that are separated from the 
vacuum by a domain wall (see \cite{Lahiri:2007ae}). In this paper we have investigated the moduli space of static
solutions of this nature. These solutions are expected to be dual to localized black holes in the 
6 dimensional bulk background. 

It may be possible - and would be very interesting - to directly construct the fluid equations that are 
dual to long distance gravitational dynamics in our background. As we have explained above, all bulk 
terms in these equations of motion are already known. However the fluid configurations in this paper 
also have boundaries which play a crucial role in their dynamics. In addition to bulk density and shear 
waves, such configurations will support fluctuations localized at boundaries; these two types of fluctuations
will interact. All of these effects (which may be important in order to study dynamical perturbations of 
the static solutions investigated in this paper) should directly follow from a thorough investigation 
of the gravitational equations.

In this paper we have also compared the phase diagram of black holes and black rings in Scherk-Schwarz 
compactified $AdS_6$ with that in asymptotically flat space. The most important difference between the two being 
the absence of black holes with spherical horizon topology in Scherk-Schwarz compactified $AdS_6$
(the one corresponding to  Myers Perry black holes in flat space) beyond a certain angular momentum.
The portion of the phase diagram which connects the ball type of solution to the ring is quite speculative in
asymptotic flat space. It could be like what we find here for asymptotically Scherk-Schwarz $AdS_6$ or something else.
Any conclusive statement about this region in six dimensional asymptotically flat space demands 
an intensive gravity calculation.

The investigation of fluctuations about our solutions may be interesting for several reasons. It is possible 
that some of the ring solutions in this paper have Plateau-Rayleigh type instabilities (the instability 
for a tube of fluid to break up into a chain of droplets) which would be dual to Gregory-Laflamme 
instabilities of the corresponding black rings 
(see \cite{Cardoso:2006sj, Cardoso:2006ks, Cardoso:2007ka, Brihaye:2007ju} for more details).
In more generality, it certainly seems interesting 
to investigate the dynamical stability of all the solutions studied in this paper. 

Finally we would like to draw attention to a curious fact that has emerged from our analysis. Within the
fluid dynamical approximation the ring solutions we have studied in this paper have the fluid rotating 
at a finite universal speed - $\frac{1}{\sqrt{6}}$ - in the extreme thin ring limit 
(this is also the upper bound on the speed of rotation).  It turns out that 
this extreme limit cannot be reliably studied within the fluid approximation.
Nonetheless the qualitative feature of an upper bound to the velocity of rotating rings has also been 
discovered in flat space \cite{Emparan:2007wm}
\footnote{we thank R. Emparan for pointing this out.},
though in that case the upper bound velocity is $\frac{1}{\sqrt{3}}$. This makes us suspect that the `speed limit'
uncovered in our analysis is not an artifact of the fluid approximation, but is likely to persist in an 
exact analysis of black rings in $AdS$ space. This is an issue that calls for further study.

\section{Acknowledgements}
The authors would like to thank Shiraz Minwalla for suggesting this problem and
providing guidance throughout the project. We  would also like to thank S. Lahiri for
helpful comments during the project and R. Emparan for insightful comments on the final draft.
We also thank all the students of Theory Physics student's room in TIFR especially 
S. Bhattacharyya, R. Loganayagam, A. Sen, B. Dasgupta, S. Ray and D. Banerjee
for help and several useful discussions.

\bibliographystyle{style}
\bibliography{bibfile}

\end{document}